\documentclass[multphys,vecphys]{svmult}

\usepackage{makeidx}         % allows index generation
\usepackage{graphicx}        % standard LaTeX graphics tool
\usepackage{multicol}        % used for the two-column index
\usepackage[bottom]{footmisc}% places footnotes at page bottom
\makeindex             % used for the subject index
\begin{document}

\title*{Young Massive Clusters - Formation Efficiencies and (Initial) Mass Functions}
\author{S{\o}ren S.\ Larsen\inst{1}}
\authorrunning{S.\ S.\ Larsen}
\titlerunning{Young Massive Clusters}
\institute{Astronomical Institute, Utrecht University, Princetonplein 5,
  NL-3584 CC, The Netherlands.
\texttt{larsen@astro.uu.nl}
}
\maketitle

\begin{abstract}
Globular clusters are often assumed to be good tracers of major star formation
episodes in their host galaxies. While observations over the past 2 decades
have confirmed the presence of young objects with globular cluster-like
properties in many galaxies, it is still not well understood exactly how the
formation efficiency of bound star clusters relative to field stars and
their 
mass spectrum depend on external factors. The cluster initial mass function 
typically 
appears to be consistent with a power-law with a slope $\sim-2$, but
most attempts
to constrain any upper limit on the CIMF have been limited by
size-of-sample effects.  
 However,
evidence is starting to accumulate for possible truncation of the cluster 
mass function.  It is tentatively suggested that the upper mass limit may 
currently be at $\sim10^5$ M$_\odot$ in the Milky Way disk, while there are
indications that it is $\sim5\times10^5$ M$_\odot$
in M51 and about $2\times10^6$ M$_\odot$ in the Antennae. Some extreme 
starbursts (e.g.\ Arp 220, NGC~7252) are (or were) able to form clusters as 
massive as $\sim10^7$ M$_\odot$.
The overall formation efficiency of star clusters (relative to field stars)
in the Galactic bulge may not have been much different from that 
in the disk today, but was probably significantly higher for
metal-poor GCs in halos.
\end{abstract}

\section{YMCs -- Guides to Young Stellar Populations?}
\label{sec:intro}

Three or four decades ago, globular cluster (GC) formation was thought by many
to be a phenomenon occurring only in the early Universe (e.g.\ \cite{pd68}).  
In the meantime, young, compact star clusters with masses in the range 
$10^4$--$10^6$ M$_{\odot}$ have been found in many different galaxies
(e.g.\ \cite{lar05,schweizer,whit01}), and there is a growing concensus that 
these ``Young Massive Clusters'' (YMCs) may well be young analogues of the 
old GCs associated almost universally with the spheroidal components of 
galaxies.  By implication, GCs have become potentially interesting not just as
fossil left-overs from the early Universe, but more generally as
test particles for studies of extragalactic stellar populations. 

In keeping with this spirit, contemporary
observing proposals or papers on GCs often include an
introductory remark along the lines of: \emph{``GCs are thought
to be good tracers of the major star forming episodes in their host
galaxies''}.  
YMCs/GCs may indeed trace star formation quite generally, but it is also 
clear that some caution must be exercised.
For example, the number of GCs per field star 
varies from galaxy to galaxy (the classical ``specific frequency 
problem''), as well as between stellar populations within galaxies 
\cite{hh02,ffg05}.  This must be due to 
differences in the formation efficiency of GCs relative to field stars, 
in the GC survival rates, or (more likely) some combination of
the two.  Of the roughly 150 catalogued GCs in our own Galaxy \cite{harris96}, 
about 2/3 are associated
chemically, spatially and kinematically with the halo, while this is true
for only $\sim1$\% of the stars. Conversely, some 90\% of the stellar
mass resides in the thin disk of our Galaxy, while no GCs are currently
known to be associated with this component.
Even though some quite massive clusters might be located in
remote parts of the Galactic disk \cite{clark05,figer06}, a simple scaling 
by stellar mass of the GCs in the bulge or halo would predict hundreds 
or thousands of 
GCs in the disk, seemingly at odds with the observations. 

All this is of course well known, and is the reason why GCs are often
assumed to trace only ``\emph{major} star forming episodes''.
The problem then remains to define when a star forming episode 
qualifies as ``major''. Perhaps GCs mainly trace spheroid formation
\cite{bs06}, but some stars
currently residing in spheroids may originally have formed in disks.
As an example, it is illustrative to consider the outcome of
merging two Milky Way galaxies: The merger product would
contain about 200 metal-poor and 100 metal-rich pre-existing GCs from the 
progenitor
galaxies, assuming no GCs are destroyed.  The current gas mass 
in the Milky Way is about $0.5\times10^{10}$ M$_{\odot}$ \cite{co96}, which 
is about half the mass of the bulge. Assuming that the merger would form GCs 
with the same efficiency as the bulge, using all the available gas, about 
50 new metal-rich GCs would form (and survive).  The resulting GC 
population would
then consist of three major sub-components: metal-poor, old GCs originating
in the progenitor galaxy halos, moderately metal-rich GCs from the
pre-existing bulges, and very metal-rich GCs formed in the merger. 
The estimate of the number of new GCs is obviously very crude. 
However, the main point here is that while the majority of the \emph{stars} 
in the resulting spheroid (about 90\%) would have formed in the disks 
already before the merger took place, their formation history might not 
be reflected in the GC system.

Mergers at higher $z$ were likely more gas-rich, and the discrepancy 
between the star- and GC formation histories may have been less extreme
than in the example outlined above. Nevertheless, since GCs play such an
important role in attempts to constrain the star formation histories of
early-type galaxies, there is a clear need to also quantify the 
limitations better.

\section{What can we learn from studies of YMCs?}

In order to understand the differences between properties of star cluster 
populations in different galaxies better, it is useful to divide the problem
into three sub-problems which can, to a large extent, be addressed
separately: 1) Understanding the \emph{cluster (initial) mass function} 
(CIMF) -- Is it universal, or do some parameters (e.g.\ the slope or upper \
mass limit) vary as a function of external parameters 
(star formation rate, gas pressure/density)?  2) The cluster \emph{formation 
efficiency} relative to ``field'' stars: Again, how does this depend on 
external factors?  3) \emph{Disruption effects:} these are driven both by 
internal mechanisms (two-body relaxation, binaries, black holes) and 
external factors (shocks, interactions with giant molecular clouds, tidal 
fields).  However, it is a hard problem and progress has been slow in making 
reliable, quantitative predictions for the time scales involved,
at least until very recently.

In the following I will concentrate mainly on the first item in the list, 
with only a few remarks about formation efficiencies. Disruption effects are 
covered elsewhere in this volume (e.g.\ Baumgardt, De Marchi, Vesperini).

\subsection{The Cluster Initial Mass Function}
\label{sec:cimf}

The number of galaxies with reliable constraints 
on the CIMF remains small. Probably the best-known
example is the Antennae, where the CIMF appears well 
approximated by a power-law $dN/dM \propto M^\alpha$ with 
$\alpha\approx-2$ over the range $10^4 < M/M_\odot < 10^6$ \cite{zf99}.
Similar mass functions have been found 
in M51 \cite{bik03}, NGC~3310 and NGC~6745 \cite{deg03}, the Milky Way 
\cite{ee97} and in the LMC \cite{hunter03}, although not all studies cover
the same mass range. It seems reasonable to conclude that
star clusters typically form with a mass spectrum that can be
well approximated by a uniform power-law over some mass range. 
It should be mentioned that some dwarf galaxies contain a few 
clusters which are much brighter than one would expect from the total
number of star clusters in those galaxies \cite{bhe02}. Here, however, I
mainly focus on the opposite problem, i.e.\ whether there might be a
\emph{truncation} of the CIMF at some upper mass $M_{\rm trunc}$, thus 
inhibiting efficient formation of clusters above a certain mass limit
($M_{\rm trunc}$) under certain circumstances (e.g.\ in the Milky Way 
disk today). 

\subsection{Some considerations on the Milky Way}
\label{sec:mw}

Starting again with the Milky Way, it is interesting to consider the
consequences of postulating that young clusters are drawn purely at random 
from a power-law distribution with $\alpha=-2$. The current star formation
rate in bound star clusters in the solar neighbourhood is estimated to
be around $5.2\times10^{-10}$ M$_\odot$ yr$^{-1}$ pc$^{-2}$ \cite{lam05}. 
Assuming for simplicity a constant cluster formation rate in the Galactic 
disk over the past 10 Gyrs and that the cluster formation rate in the
Solar neighbourhood is representative for the disk as a whole, this 
corresponds to $\sim10^9$ M$_\odot$ formed in bound clusters within the 
Solar circle (8.5 kpc). This is most likely a conservative estimate and
the true number may well be significantly higher.
Sampling these clusters from 
a power-law CIMF for 
$10^2 < M/M_\odot < 10^7$, one predicts a total of close to $9\times10^5$
clusters formed over the lifetime of the disk, of which 9000, 900 
and 80 have $M>10^4 M_\odot$, $M > 10^5 M_\odot$ and $M > 10^6 M_\odot$. 
These numbers do not depend strongly on the adopted upper and lower 
integration limits, although they do depend on the CIMF slope -- a steeper
slope implies a more bottom-heavy CIMF, with fewer high-mass clusters.

There are two consistency checks worth making: first, the stellar mass
of the Milky Way bulge is about 10\% of that of the
disk. If clusters formed with the same efficiency in the bulge
as they do in the disk now, one might expect about 90 clusters with
$M > 10^5 M_\odot$ in the bulge, of which 8 have $M > 10^6 M_\odot$.
The actual observed numbers of GCs in the bulge are smaller by a factor 
of 4, which may be partly due to disruption effects. However, there is no 
indication that clusters formed with a \emph{higher} efficiency in the 
bulge, and the number of GCs observed in the bulge may be roughly 
consistent with a formation efficiency (relative to field stars) similar 
to that observed today in the Galactic disk.  The numbers for the
halo are more difficult to reconcile with this picture, since the halo
has about twice as many GCs as the bulge but an order of magnitude fewer
stars. This suggests a higher formation efficiency of metal-poor halo GCs, 
consistent with observations of early-type galaxies \cite{hh02,ffg05}.  The 
distinction between \emph{halo} (metal-poor) GCs and all other star clusters 
may be more fundamental than the one between old GCs in general and 
present-day star formation. Note that these arguments are slightly different 
from those of McLaughlin \cite{mc99} who argued for a universal cluster 
formation efficiency relative to the total available \emph{gas} mass.

Second, we can compare with the number of massive star clusters actually 
observed in the disk. By construction, the current formation rate agrees
with the number of low-mass clusters ($M<10^3$ M$_\odot$) observed locally,
but it is interesting to see what happens when extrapolating to higher
masses.  Recently, at least two young clusters with masses in the range 
$10^4$--$10^5$ M$_\odot$ and ages $<10^7$ years have been identified 
\cite{clark05,figer06}. If they are taken as representative of the formation 
rate of such objects within a distance of 5 kpc, this corresponds to 
2 kpc$^{-2}$ Gyr$^{-1}$, or 4500 such clusters formed within the Solar circle 
over 10 Gyrs. Again, this is roughly consistent with the order-of-magnitude 
estimates above (in fact, 8000 clusters with $10^4<M/M_\odot<10^5$ are 
predicted), suggesting that clusters like Westerlund 1 with masses up to 
about $10^5$ M$_\odot$ occur naturally (albeit rarely) as part of the normal 
hierarchy of star formation in the Milky Way disk today.

%  When relating the observed number of clusters to the formation rate,
%we must also consider disruption effects.  
A typical disruption time for a cluster with mass $M$ in the solar 
neighbourhood is $t_{\rm dis} = 1.3 \, {\rm Gyrs} \, (M/10^4 M_\odot)^{0.62}$ 
\cite{lam05}.  Assuming that this scaling remains valid for $M>10^4$ 
M$_\odot$, a cluster with an initial mass of $10^5$ M$_\odot$ is expected to
disrupt in about 5 Gyrs while a $10^6$ M$_\odot$ cluster has
a lifetime well in excess of a Hubble time. The disk should then
contain about 500 clusters with masses greater than $10^5$ M$_\odot$
and still virtually all of the 80 clusters with $M > 10^6$ M$_\odot$ formed 
over its lifetime.  Of these objects, 7 should be found within a distance 
of 1 kpc. This estimate makes the crude assumption that clusters disrupt
instantaneously, while in practice mass is lost at a nearly constant rate
over the lifetime of the cluster. 
Although current catalogs of Milky Way open clusters are highly 
incomplete beyond 1 kpc, it seems unlikely that a large population
of clusters with masses in the range $10^5$--$10^6$ M$_\odot$ could have 
been missed.  It appears plausible that the CIMF in the Milky Way 
is truncated somewhere in the vicinity of $10^5$ M$_\odot$, or at
least becomes much steeper than a power-law with a slope of $-2$.
However, it would be highly desirable to quantify better the
completeness of current cluster surveys in the disk out to large
distances (several kpc).

\subsection{Constraints on the CIMF in other galaxies}

Studies of the CIMF in external galaxies are complicated by the rapid
change in mass-to-light ratio that characterizes simple stellar 
populations. Observed
luminosities cannot be converted to masses unless the age of each individual
cluster is known with some accuracy, which generally requires U-band 
imaging. Several studies have shown that the \emph{luminosity} function of
young star clusters generally appears to be sampled all the way up to its 
statistical upper limit \cite{whit01,bhe02,lar02}.  Even the scatter around 
the predicted relation 
is consistent with random sampling \cite{lar02}. No direct evidence for 
truncation of the LF has been found so far, suggesting that most large 
galaxies are physically able to form star clusters with masses 
\emph{at least} up to about $10^5$ M$_\odot$.

\begin{figure}
\centering
\includegraphics[width=11cm]{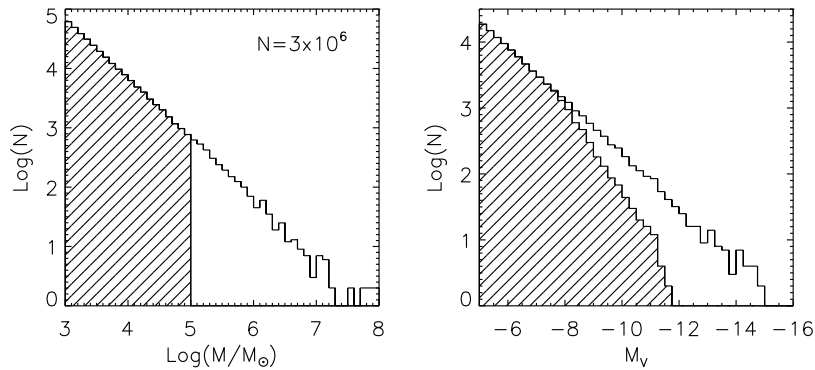}
\caption{Mass and $M_V$ distribution for a simulated cluster population
 with $3\times10^6$ clusters and 6$<$log(age/yr)$<$9.
 Hashed and outlined histograms are for mass distributions truncated 
 at $10^5$ M$_\odot$ and $10^8$ M$_\odot$.}
\label{fig:histo}       % Give a unique label
\end{figure}

In general, there is no straight-forward way to infer the mass
function of a cluster sample from the observed luminosity function
\cite{zf99}. Only
in the special case where the MF is a simple, untruncated
power-law, or the age distribution is a delta function, will the LF and MF
have the same shape. This point
is illustrated in Fig.~\ref{fig:histo}, which compares the mass- and
luminosity functions for simulated cluster samples with power-law
mass functions truncated at $10^8$ M$_\odot$ and $10^5$ M$_\odot$.
The clusters were assigned random ages uniformly distributed between 
$10^6$ years
and $10^9$ years and masses were converted to $M_V$ magnitudes using
SSP models \cite{bc03}. Mass loss and cluster disruption were ignored.
In the left-hand panel, the MF truncation occurs as a simple cut-off, while 
in the LF (right) the mass cut results in a steepening of
the slope at the bright end rather than any distinct cut-off in luminosity. 
For a real cluster population, the maximum mass may be inferred from
the ``bend'' that occurs at $M_V\approx-8$ in Fig.~\ref{fig:histo}, assuming
that the MF is a power-law up to $M_{\rm trunc}$
\cite{gieles06}.  Such
a bend has been observed in the Antennae and M51, where it may be
explained by a MF truncated near $2\times10^6$ M$_\odot$ and
$\sim5\times10^5$ M$_\odot$ \cite{zf99,gieles06}. However, in still more 
active
galaxies such as Arp 220, NGC~7252 and NGC~1316, there are clusters with 
masses as high as $\sim10^7$ M$_\odot$ \cite{bastian06,wilson06}.

Let us now consider the behaviour of the following quantities as a function
of the total number of clusters ($N$) in a population with a 
truncation at an upper mass limit $M_{\rm trunc}$: 1) the maximum mass
$M_{\rm max}$ occurring in the population, 2) the mass of the brightest
cluster $M_{\rm brightest}$, and 3) the magnitude of the brightest
cluster, $M_V^{\rm brightest}$.
If $N$ clusters are sampled at random from a power-law $dN/dM\propto M^{-2}$
with lower mass limit $M_{\rm min}$, then statistically the most massive
cluster will have $M_{\rm max} = N \, M_{\rm min}$ \cite{whit01,bhe02}. 
From Monte-Carlo simulations of 
various cluster formation histories, Weidner et al.\ \cite{wkl04} found that 
when clusters are sampled at random from a power-law MF, the most massive 
cluster is also the brightest in about 95\% of the cases. If the mass
function has a real physical upper limit, this is not necessarily the
case. 

\begin{figure}
\noindent \begin{minipage}{121mm}
\includegraphics[width=5.5cm]{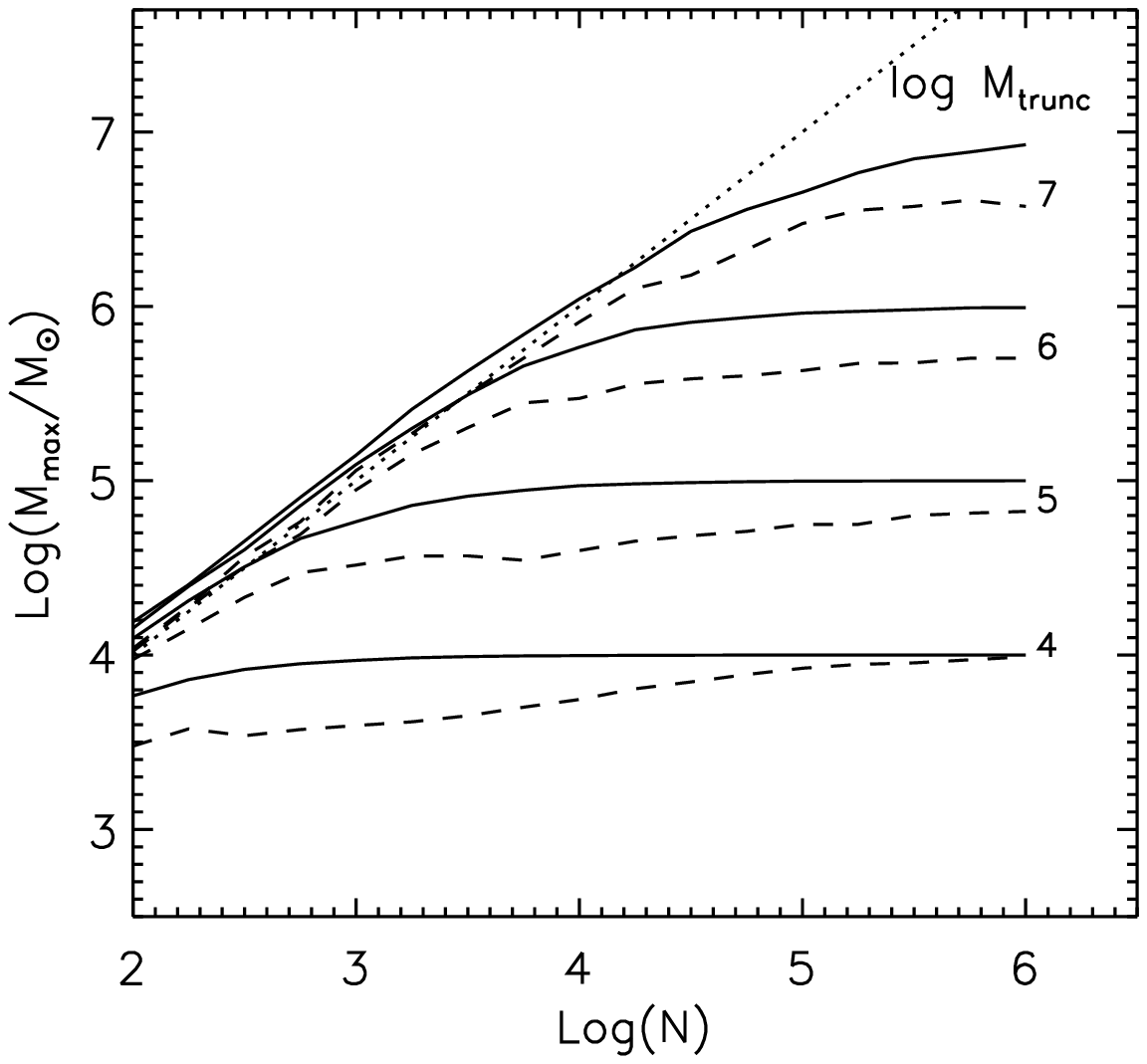}
\includegraphics[width=5.5cm]{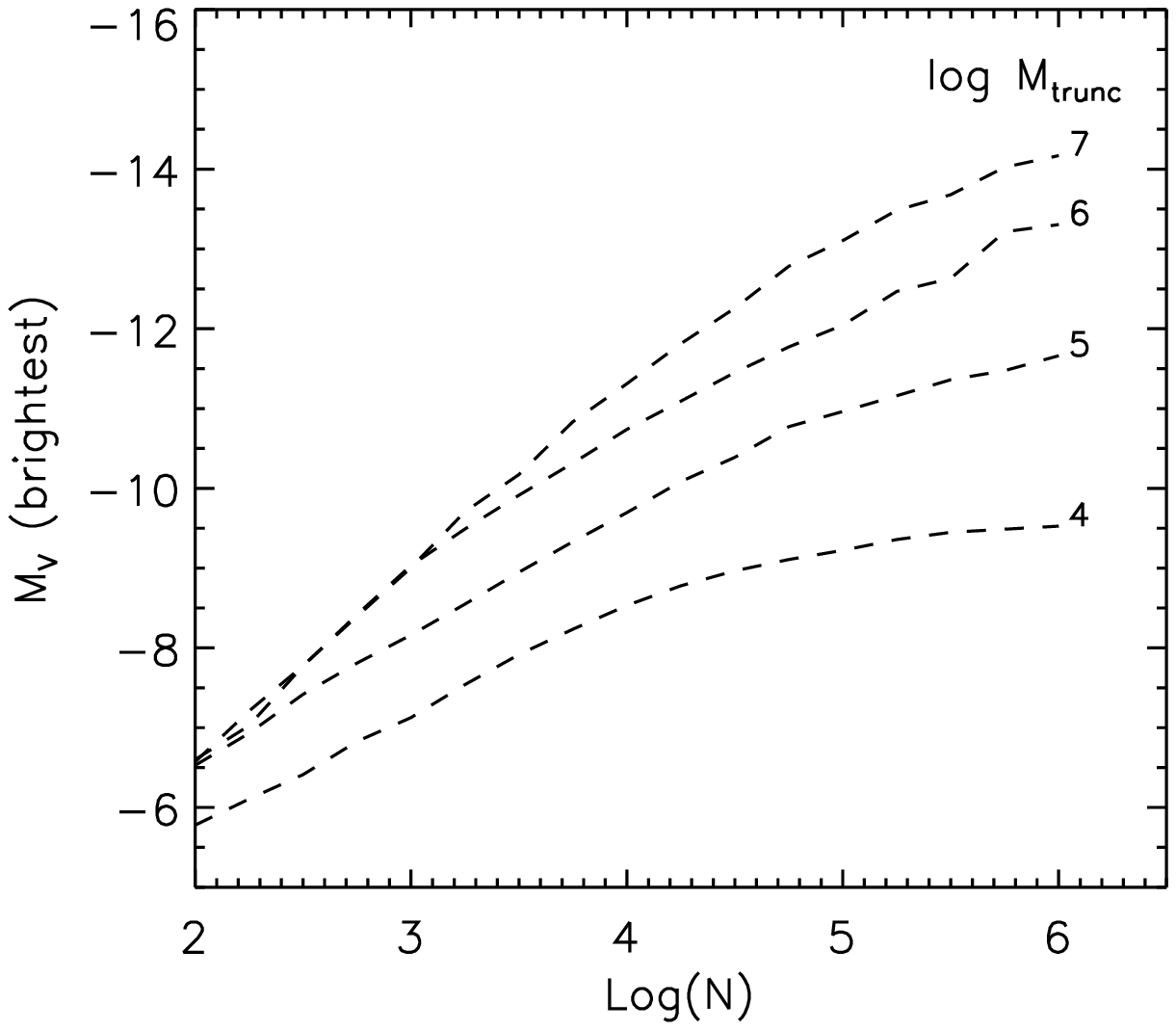}
\end{minipage}
\caption{Left: Mass of the most massive (solid lines) and most luminous 
(dashed lines) cluster as a function of total number of clusters in simulated 
cluster samples with various fixed upper mass limits. The dotted line is
the maximum cluster mass expected from random sampling from an
untruncated power-law. Right: The luminosity of the brightest cluster as
a function of $N$ and $M_{\rm trunc}$.}
\label{fig:mvn}       % Give a unique label
\end{figure}

The left panel in Fig.~\ref{fig:mvn} shows the results of Monte-Carlo 
simulations
for $M_{\rm max}$ and 
$M_{\rm brightest}$ in cluster samples with 
various $M_{\rm trunc}$ limits.  Clusters were drawn at
random from the same population used in Fig.~\ref{fig:histo}, truncated
at $M_{\rm trunc}$ values between $10^4$ M$_{\odot}$ 
and $10^7$ M$_{\odot}$. 
The median values of $M_{\rm max}$ and $M_{\rm brightest}$ in 1000
experiments are shown with
solid and dashed lines as 
a function of the number of clusters sampled. The relation predicted 
by pure sampling statistics is indicated as a dotted line. For small $N$ and 
large $M_{\rm trunc}$, the truncation of the MF is not ``felt'' and the
$M_{\rm max}$ and $M_{\rm brightest}$ vs.\ $N$ relations approach the
curve for random sampling, i.e.\ the Weidner et al.\ result is reproduced.
Conversely, for small $M_{\rm trunc}$ and large $N$, statistical effects
become unimportant and $M_{\rm brightest} \, \sim \, M_{\rm max} \, \sim \,
M_{\rm trunc}$. In between these extremes there is a regime
where the mass function is likely to be sampled up to higher masses than
the luminosity function, or $M_{\rm max} > M_{\rm brightest}$. 
This turns out to be the situation encountered in many real galaxies.

For both $M_{\rm max}$ and $M_{\rm brightest}$, Fig.~\ref{fig:mvn} shows a 
fairly rapid transition between the regimes where size-of-sample effects and 
truncation dominate. Thus, it might seem that the observed strong relation
between total number of clusters in galaxies and the luminosity of the
brightest cluster \cite{whit01,bhe02,lar02} is incompatible with 
truncation of the CIMF playing any important role. However, as shown in
the right-hand panel of Fig.~\ref{fig:mvn}, $M_V^{\rm brightest}$ has a
steep dependency on $N$ over a much wider range in $N$ than 
$M_{\rm max}$ and $M_{\rm brightest}$. This is because the mean \emph{age}
of the brightest cluster shifts towards younger ages for higher $N$,
as it becomes increasingly likely to encounter a cluster with mass
near $M_{\rm trunc}$ in the brief phase where the M/L ratio is very low.

From the preceding discussion it should be clear that the LF can be dominated 
by sampling effects, even if the MF does have a physical truncation. For 
example, for $M_{\rm trunc}=10^5$ M$_{\odot}$, the $M_{\rm max}$ curve starts 
to flatten at $N\sim10^3$, implying that the MF would be sampled up to its 
physical upper limit already in a relatively cluster-poor galaxy.  The 
$M_V^{\rm brightest}$ curve, on the other hand, would continue to rise 
beyond $N=10^5$, corresponding roughly to a Milky Way-sized galaxy 
(cf.\ Section~\ref{sec:mw}). 

The key to detecting a physical
upper limit of the CIMF will be to cover a dynamic range large enough to 
study the LF in
detail and detect signatures such as 
the ``bend'' in Fig.~\ref{fig:histo}. Ultimately, however, inferences about
the MF based on the LF remain indirect and dependent on assumptions about
the shape of the MF, star formation history etc., and ideally it would be
desirable to have direct information about the mass functions in several
more galaxies. Several such studies are now underway, and may provide
important insight into the MF in the not too distant future.

\section{Concluding remarks}

While research in extragalactic star clusters remains a very active field, 
we are still facing a 
number of important questions. The ubiquity of YMCs in external galaxies is 
now well established, but the apparent absence of a large population of 
clusters with masses in the range $10^5$--$10^6$ M$_\odot$ in the Galactic 
disk remains somewhat of a puzzle. This may suggest an upper limit to the 
CIMF near $10^5$ M$_\odot$ in the Milky Way, as noted already by van den 
Bergh \& Lafontaine \cite{vl84}, although studies of disk clusters are still 
hampered by our own location close to the Galactic plane.  However, the
recent realization that the Milky Way is still forming clusters with 
masses near $10^5$ M$_\odot$ hints that the CIMF in the Milky Way
may not be very different from that in the LMC, the main difference being 
that the census of YMCs is more complete
in the LMC.  There is some evidence for truncation 
at a somewhat higher mass in M51 ($\sim5\times10^5$ M$_\odot$) and the 
Antennae ($\sim2\times10^6$ M$_\odot$), while Arp 220, NGC~1316 and NGC~7252 
host clusters as massive as $\sim10^7$ M$_\odot$. These galaxies also define
a sequence of increasing star formation rate, suggesting that galaxies
with higher SFRs are physically able to form more massive clusters. This
may provide a hint as to why GC formation was common at high $z$, when 
SFRs and gas densities were generally higher.

\printindex
\end{document}